\documentclass[prl,twocolumn,showpacs,superscriptaddress]{revtex4}
\usepackage{graphicx}
\usepackage{color}
\usepackage{mathrsfs}

\begin{document}

\title{X-ray Coherent diffraction interpreted through the fractional Fourier transform}
\author{D. Le Bolloc'h},\author{ J.F. Sadoc}
\affiliation{Laboratoire de Physique des Solides (CNRS-UMR 8502), B{\^a}t. 510, Universit{\'e} Paris-sud, 91405 Orsay cedex, France}

\begin{abstract}
Diffraction of coherent x-ray beams is treated through the Fractional Fourier transform. The transformation allow us to deal with coherent diffraction experiments from the Fresnel to the Fraunhofer regime. The analogy with the Huygens-Fresnel theory is first discussed and a generalized uncertainty principle is introduced.
\end{abstract}

\maketitle

%------------------------------------------------------------------------------
\section{Diffraction of a rectangular aperture}

Almost two centuries ago, Fresnel has shown a beautiful consequence of the wave character of light. If an opaque disc is located in front of a monochromatic light source, a bright spot is observed at a given distance $z_b$ at the center of the shadow of the disc\cite{fresnel}. This phenomenon is known as the {\it Poisson bright spot}. In the opposite situation, the diffraction pattern of an aperture gives rise to a minimum of intensity if the detector is located at a very specific distance. This minimum of intensity will be called the $dark$ spot in the following.

\begin{figure}[!ht]
 $$\resizebox{1\columnwidth}{!}{%
\includegraphics{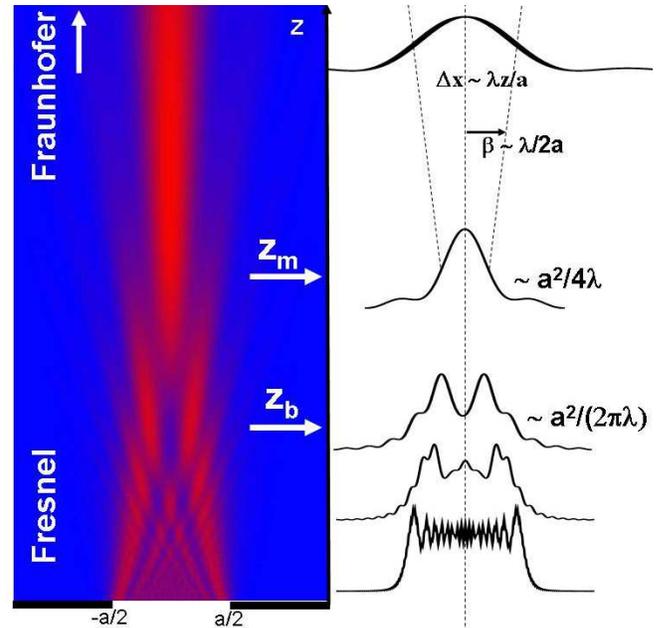}
}$$
\caption{Diffraction of a rectangular aperture in the Fresnel regime and corresponding profiles simulated from the Fresnel integral (eq. \ref{difAmp}) or from the Fractional Fouier Transform (eq. \ref{frftPF}).
The dark spot is located at $z_b$ and the minimum beam width at $z_m$.
\label{fresnel_fraunhofer}}
\end{figure}

The dark spot results from a destructive interference. It is located in the Fresnel regime where the beam remains almost parallel and where the amplitude oscillates rapidly. 
The diffraction pattern of a squared aperture simulated from the Fresnel's integral is displayed in Fig~\ref{fresnel_fraunhofer} where the dark spot is clearly visible.

The Fraunhofer regime is reached for  distances  $z$, between the detector and the diffracted object\cite{wolf}, larger than
$
z>>\frac{2\pi a^2}{\lambda}.
$
In this regime, the diffracted amplitude is proportional to the Fourier Transform (FT) of the amplitude at the diffracted object position:
\begin{equation}
F(q)=\frac{1}{\sqrt{2\pi}}\int_{-\infty}^{\infty} f(x) e^{-i q x} dx,
\label{ft}
\end{equation}
with $q=\frac{2\pi}{\lambda}$.
In the case of a rectangular
function s(x) which is only non-zero in the interval [$-\frac a2$,$\frac a2$], we obtain:
\begin{equation}
I(q)=F^*(q)F(q)\propto [\frac{\sin(q\frac a2)}{q \frac a2}]^2.
\end{equation}
 The diffracted intensity by a rectangular function is proportional to the well-known cardinal sinus function squared,
in perfect agreement with measurements. The measurement of the diffraction pattern of a 2$\mu m\times$2$\mu m$ slit by a x-ray beam in the Fraunhofer regime is displayed in Fig~2. In this regime, the amplitude
varies very smoothly and the beam width $\Delta$ is proportional to $z$:  $\Delta\propto \lambda z/a$.

This old problem of optic has recently recovers a great interest by the scientific community using coherent X-ray diffraction. Indeed, coherent X-ray beams are obtained from weakly coherent synchrotron sources by cleaning and collimating the beam thanks to rectangular slits\cite{lebolloch}. Concretely, two sets of rectangular slits are usually used along the optical path. The downstream one is located at few centimeters from the sample and is opened at few tens of micrometers. Within this setup, the knowledge of the dark spot position is important because it is usually located at few tens of centimeters downstream the slit. As a consequence, the sample may be located either in the Fresnel or in the Fraunhofer regime  as respect to aperture and wavelength, which may strongly influence diffracted patterns.

\begin{figure}[!ht]
$$\resizebox{0.35\textwidth}{!}{%
 \includegraphics{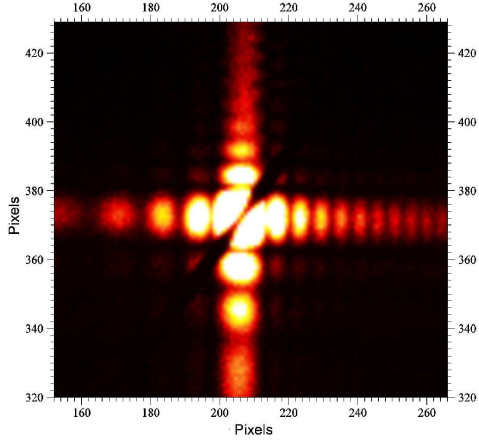}
}$$
\caption{Diffraction pattern of a $2\mu m\times2\mu m$ squared aperture obtained at $\lambda=1.5 \AA$ (Log scale). The camera is located 2,22m downstream. A diagonally placed absorbing wire has been used across the direct beam to protect the camera (from reference \cite{jacques}).
\label{direct_centre}}
\end{figure}

The dark spot position is not obvious to calculate analytically.
In first approximation, $z_m$  can be estimated by assuming that it corresponds to the distance $z$ where the beam width in the Fraunhofer regime ($\Delta\approx\lambda \frac za$) is equal to the beam width in the Fresnel regime ($\Delta\approx a$), that is:
$$z_m=\frac{a^2}{\lambda}.$$
From numerical simulations, the dark spot position is located close to $z_b\approx\frac{ a^2}{2\pi\lambda}$ and the location where the beam width is minimum at $z_m\approx\frac {a^2}{4\lambda}$ (see Fig~\ref{fresnel_fraunhofer}).

 \section{Fractional Fourier Transform}
 
 We first discuss in this section the relation between the Huygens-Fresnel theory and the fractional Fourier transform,
 introduced by Namias\cite{namias}. The previous slit diffraction in Fig~\ref{fresnel_fraunhofer} can be  obtained from the fractional Fourier Transform. Within this framework, the resolution of the Fresnel's integral
 is nothing else but the resolution of the quantum harmonic oscillator. 

 \subsection{Fractional Fourier Transform: an operator}  

\begin{figure}[!ht]
$$\resizebox{0.5\textwidth}{!}{%
 \includegraphics{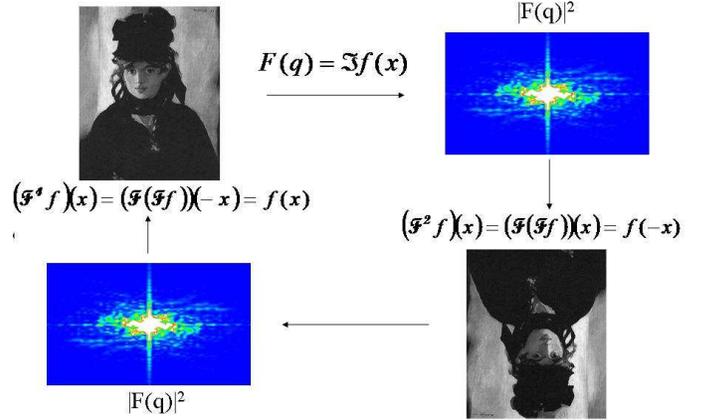}
}$$
\caption{A classical Fourier transform is an angular transformation of  $\pi/2$. The operator $\mathcal{F}$ has to be applied four times to recover the original portrait. The starting point of the simulation is a painting by Manet "Berthe Morisot" Mus\'ee d'Orsay, Paris (displayed in black and white and considering brightness variations as density variations). Only the intensity is represented in the reciprocal space.
\label{fractional}}
\end{figure}

 The most natural way to introduce the Fractional Fourier Transform is to note that  the Fourier Transform,
 defined in eq.\ref{ft}, can be written as an operator $\mathcal{F}$ acting on function: 
 $$
 F= \mathcal{F} [f].  
 $$
If the same operator $\mathcal{F}$ is applied two times, we obtain:
 $$
 \mathcal{F}^2[f](x)=f(-x).
 $$
 The operator $\mathcal{F}$ has to be applied four times to recover the original function:
 $$
 \mathcal{F}^4[f](x)=f(x)
 $$
 as illustrated in Fig~\ref{fractional} from a 2D picture.
Within this framework, the standard  Fourier Transform (FT) is an angular transformation of  $\pi/2$  in the (x,q) plane.
 A generalized Fourier Transform can thus be developed for any angle of rotation $\alpha$ from zero to $\frac \pi 2$. 
 The integral form of the fractional Fourier transform (in 1D case) can be written as\cite{namias,bultheel}:
 \begin{equation}
 \mathcal{F}^\alpha [f](x)=\int_{-\infty}^\infty K(x,u) f(u) du
 \label{frft}
 \end{equation}
 with
 $$
 K(x,u)=\sqrt{\frac{1-i\cot\alpha}{2\pi}}\exp[{-i\frac{xu}{\sin\alpha}+\frac i2(x^2+u^2)\cot\alpha}]
 $$
 For $\alpha=0$, it can be shown that this expression is equivalent to the identity\cite{mcbride}. The standard
 Fourier transform (eq. \ref{ft}) is easily obtained for $\alpha=\pi/2$.

\subsection{The fractional Fourier transform and slit diffraction}

%Modification JFS%%%%%%%%%%%%%%%%%%%%%%%%%%%%%%%%%%%%%%%%%%%
The Fractional Fourier Transform (FrFT) in 2D case, so with appropriate factor in front of integral is:
\begin{eqnarray}
\mathcal{F}_{\alpha} f({\bf q})= \frac{-i\alpha}{2\pi \sin \alpha} \exp(-\frac{i}{2} {\bf q}^2 \cot \alpha) \dots \nonumber\\
\dots \int_{R^2} \exp(-\frac{i}{2} {\bf r}^2\cot \alpha) \exp(-\frac{i {\bf q}.{\bf r}}{\sin \alpha} ) f({\bf r})d{\bf r}
\label{frftPF}
\end{eqnarray} 

It has to be compared to the diffracted field amplitude. We consider a diffracting plane $\Sigma$ (containing the slit) and we observe the field amplitude at distance $z$ of $\Sigma$ on a plane screen $\Pi$ orthogonal to the axis. The field amplitude at a position on $\Pi$ defined by the variable $\xi$ is:
\begin{eqnarray}
A_{\Pi}(q)= \frac{1}{\lambda z} \exp(-\frac{i \pi  {\bf \xi}^2}{\lambda z}) \dots \nonumber\\
\dots \int_{R^2} \exp(-\frac{i \pi}{\lambda z} {\bf r}^2) \exp(- \frac{2 i \pi {\bf \xi}.{\bf r}}{\lambda z} ) A_{\Sigma}({\bf r})d{\bf r} .
\label{difAmp}
\end{eqnarray} 

The similarity between equation \ref{frftPF} and    \ref{difAmp} suggests to write diffraction in terms of FrFT. Papers by Pellat-Finet \cite{pellatOpLet,pellat} give the relation between the two expressions.
The method is based on the use of intermediate spherical surface on which the amplitude of the field is considered, then within conditions on curvature and position of these surfaces,   field on it can be related by Fourier transform or fractional Fourier transform.

\subsubsection{The Fraunhofer regime}
\begin{figure}[!ht]
$$\resizebox{0.45\textwidth}{!}{%
 \includegraphics{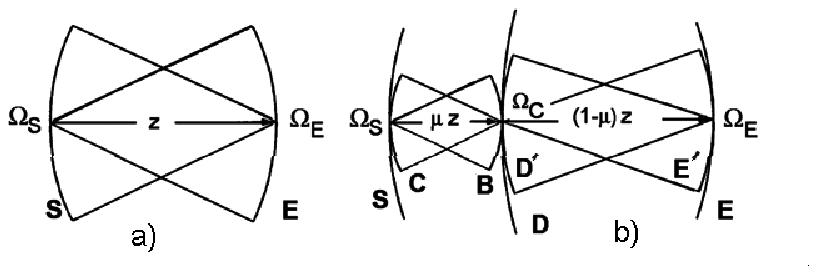}
}$$
\caption{Fraunhofer diffraction (a): the field on E is the Fourier transform of the field on S. Fresnel and fractional Fourier transform (b): the field on S is change into the field on D by a fractional Fourier transform defined by $\alpha$ with $\cot(\alpha)=(1-\mu)/\mu$. The field on S can be transform into the field on E combining two fractional Fourier transform defined by angles $\alpha$ and $\beta$ such that $\alpha+\beta=\pi/2$.}
\label{sphericalsurf}
\end{figure}

The simplest example is the case of the Fraunhofer regime. Consider two spherical surfaces (see Fig~\ref{sphericalsurf}a). The first one is the object surface $S$ of radius $z$ passing through the origin $\Omega_S$ (the middle  of the slit if the object is a slit) and with a center $\Omega_E$ on the optical axis at positive distance $z$ of $\Omega_S$. The other spherical surface $E$, a screen where the field is observed, contains $\Omega_E$ and has center $\Omega_S$, so his radius is  $-z$ if oriented from surface to center along the axis. Points on these surfaces are defined by the coordinates of their projection parallel to the axis on the planes orthogonal to the axis in $\Omega_S$ and $\Omega_E$.
In this case the field $U_S$ on $S$ and  $U_E$ on $E$ are related by a Fourier transform:
  \begin{equation}
U_E({\bf r}^\prime)= \frac{i }{\lambda z}\int_{R^2} \exp( \frac{2 i \pi }{\lambda z}{\bf r}.{\bf r}^\prime) U_S({\bf r} ) dr
\label{ TFfield}
\end{equation}
 Using reduced variables
$ \rho = (\frac{2\pi}{\lambda z})^{1/2}  {\bf r}  $ and $ \rho^\prime = (\frac{2\pi}{\lambda z})^{1/2}  {\bf r}^\prime $ as arguments of scaled functions $V_S(\rho) $ and $ V_E(\rho^\prime)$:
  \begin{equation}
V_E(\rho^\prime)= i \mathcal{F}_{\pi/2}[V_S](\rho^\prime)
\label{ TFfield2}
\end{equation}
The field on $E$ is the Fourier transform of the field on $S$. If $z\rightarrow\infty$ this gives the limit of the Fraunhofer regime: a plane object is Fourier transform into a plane diffraction at infinity. Spherical surfaces in place of planes have to be considered for finite $z$. The Fraunhofer regime corresponds to the large $z$ domain in which phases shift between planes and spherical surfaces remain small.

\subsubsection{The Fresnel regime}
In the FrFT, $\alpha=0$ corresponds to $z=0$ and $\alpha=\pi/2$ corresponds to $z\rightarrow\infty$. It is tempting to get a relationship between $z$ and $\alpha$.
Comparison of two terms $\exp(-\frac{i {\bf q}.{\bf r}}{\sin \alpha} )$ in equation \ref{frftPF} and
$\exp(- \frac{2 i \pi {\bf \xi}.{\bf r}}{\lambda z} )$ in equation~\ref{difAmp} suggests
\begin{equation}
\sin(\alpha)\propto\frac{\lambda z}{2\pi},
\end{equation}
but other terms in these equations contradict this expression mainly for small $\alpha$, in the Fresnel regime.
There is a difficulty comparing variable $\bf q$ which defines a wave vector in the reciprocal space with $\bf \xi$ in real space.

The solution to this problem has been given by Pellat-Finet. In place of the  spherical surface through $\Omega_S$ with radius $z$ a surface $C$ with a smaller radius $\mu z$ also through $\Omega_S$ is considered. Call $\Omega_C$ its center. We consider also a spherical surface $B$ through $\Omega_C$ and center $\Omega_S$, so with oriented radius $-\mu z$. Field transfer from $S$ to $C$ implies a quadratic phase factor, and transfer from $C$ to $B$ is again a Fourier transform. Consider a reduced field function $V_B$ of a reduced variable $\rho^{\prime\prime}$,
\begin{equation}
V_B(\rho^{\prime\prime})=U_B((\frac{\lambda z}{2 \pi})^{1/2}\frac{\rho^{\prime\prime}}{\cos(\alpha)+\sin(\alpha)})
\end{equation}
The reduced variable are defined by $ \rho = (\frac{2\pi}{\lambda z})^{1/2}  {\bf r}  $ and
$ \rho^{\prime\prime} = (\frac{2\pi}{\lambda z})^{1/2}  {\bf r}^{\prime\prime}  $.
Notice that variables ${\bf r}$ and ${\bf r}^{\prime\prime}$ give position of   points on   spherical surfaces $C$ and $B$. Such points are defined by coordinates of their orthogonal projection   on a plane orthogonal to the optical axis. 

The $\mu$ parameter smaller than one, is related to $\alpha$ by
\begin{equation}
\cot(\alpha)=\frac{1-\mu}{\mu} \; {\rm or} \; \mu=\frac{\sin(\alpha)}{\cos(\alpha)+\sin(\alpha)}.
\end{equation}
In order to have a fractional Fourier transform with the correct quadratic phase factor, an other spherical surface $D$ is needed. A sphere through $\Omega_C$ with positive oriented radius $R$ such that $R=(\mu^2+(1-\mu)^2)z/(1-2\mu)$. Note that surface D corresponds to a plane when $\mu=\frac 12$. Then we have:
\begin{equation}
V_D(\rho^{\prime\prime})=e^{i\alpha}(\cos(\alpha)+\sin(\alpha))\mathcal{F}_\alpha[V_S](\rho^{\prime\prime}).
\end{equation}
It can be shown that transfer from the field on $D$ to the field on $E$, using intermediated spherical surfaces $D^\prime$ and $E^\prime$ to have correct phase factors, is given by a fractional Fourier transform $\mathcal{F}_\beta$ where $\beta$  defined by the coefficient $\mu^\prime=1-\mu$  is $\beta=\pi/2 -\alpha$.

This prove the continuity of the transformation following that of the fractional Fourier transform, $\mathcal{F}_{\pi/2}=\mathcal{F}_{\alpha}+\mathcal{F}_{\beta}$. Notice that this continuity is false for the Fresnel transformation.

%%%%%%%%%%%%%%%%%%%%%%%%%%%%%%%%%%%%%%%%%%%%%%%%%%%%%
\subsection{Relation with the quantum harmonic oscillator}

 The most remarkable property of the fractional Fourier transform is that a possible choice for the eigenfunctions of the operator $\mathcal{F}_\alpha$
is given by the set of normalized Hermite-Gauss functions\cite{namias,mccellan,dattoli}, similar to the orthogonal harmonic oscillator basis:
$$\Phi_n(x) =
\frac{1}{\pi^\frac 14\sqrt{2^n n!}}\exp[-x^2/2] H_n(x),$$
where $H_n(x) = (-1)^n\exp[x^2/2]\frac {d^n}{dx}\exp[-x^{2}/2].$
The eigenvalues of $\mathcal{F}_\alpha$ are $e^{i n\alpha}$:
$$
\mathcal{F}_\alpha\Phi_n(x)=e^{i n\alpha}\Phi_n(x).
$$
Extension to two dimensions is straight forward.
Obviously, $
\mathcal{F}_{\alpha}\mathcal{F}_{\beta}\Phi_n(x)=\mathcal{F}_{\alpha+\beta}\Phi_n(x).$

Consequently the diffraction problem can be mapped on that of the harmonic oscillator.
The Fresnel's integral or the fractional Fourier transform applied to the diffraction of a squared aperture is 
equivalent to the resolution of the Schrodinger equation with time  in a harmonic potential, with  a rectangular function as initial time condition\cite{cohen}. Then time evolution is related to the $\alpha$ variation. 
In terms of quantum mechanics, the fractional Fourier transform offers the way to continuously switch from the
position space to the impulsion space.

 \subsubsection{Gaussian beams} 

\begin{figure}[!ht]
$$\resizebox{0.45\textwidth}{!}{%
\includegraphics{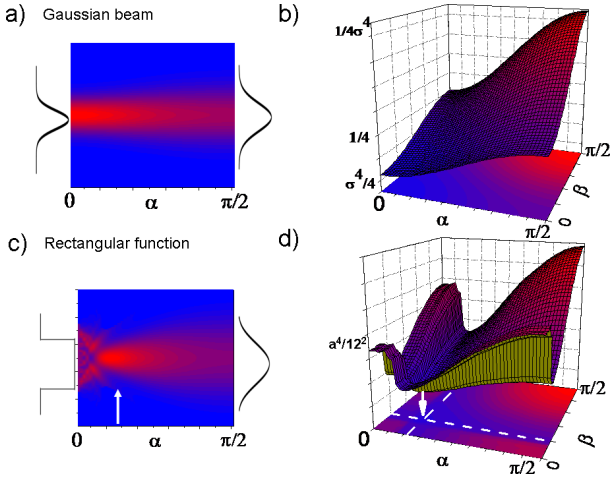}
}$$
\caption{a) Propagation of a Gaussian beam treated by the FrFT for $\alpha\in[0,\frac \pi 2]$. b) Product of variances ${\rm Var}[\mathcal{F}_{\alpha}f(x)]\times {\rm Var}[\mathcal{F}_{\beta}f(x)]$ with ($\alpha,\beta)\in[0,\frac \pi 2]$.  c) Coherent diffraction of a rectangular function treated  by the FrFT  as respect to $\alpha\in[0,\frac \pi 2]$. The minimum of the rms variance is indicated by an arrow. d) Product ${\rm Var}[\mathcal{F}_{\alpha}f(x)]\times {\rm Var}[\mathcal{F}_{\beta}f(x)]$ with $(\alpha,\beta)\in[0,\frac \pi 2]$  if the variance is measured over a limited area centered at the beam maximum. The minimum product value is indicated by an arrow.}
\label{uncertainty}
\end{figure}

 It is interesting to consider the behavior of a Gaussian beam propagating from the Fresnel to the Fraunhofer regime (see Fig~\ref{uncertainty}a).
Since the first eigenfunction of the basis is a Gaussian, it could be expected that a Gaussian beam
will remain Gaussian if its width correspond to that of this eigenfunction. Following the Pellat-Finet approach, we have introduced reduced variables $ \rho = (\frac{2\pi}{\lambda z})^{1/2}  {\bf r}  $. It is with such variables that the width of a beam have to be given. Then we have to consider, as given in Fig~\ref{sphericalsurf}, a Gaussian beam modulating a spherical wave $S$ with center $\Omega_E$ at distance $z$. If the Gaussian amplitude on $S$ is proportional to $\exp({-\frac{\rho^2}{2\sigma^2}})$ with $\sigma=1$ like for the first eigenvalue of the harmonic oscillator, this width will be kept propagating from $S$ to $E$ through $D$. Expressed using $r,r^\prime$ or $r^{\prime \prime}$ variables the width is $(\frac{\lambda z}{2\pi})^{1/2}$. With numerical values $\lambda=1\AA$ and $z=1$ m, the constant width is $4$ $\mu$m. The physical meaning of this value is that a beam ($\lambda=1\AA$) falling on a curved mirror designed to focused at $1m$, with a width of $1\mu m$ just after the mirror, have in fact a constant width.

What is the behavior of narrower or wider beams ($\sigma\neq$1) than the first eigenfunction  ? The  1-D fractional Fourier transform $\mathcal{F}^{\alpha}f(x)$ of a Gaussian function $f(x)$,
\begin{equation}
f(x)=\frac{1}{\pi^{\frac 14}\sqrt\sigma}\exp({-\frac{x^2}{2\sigma^2}}),
\label{gaussianbeam}
\end{equation}
is:
\begin{equation}
F_\alpha(q)=\frac{\sqrt{\sigma} \sqrt{\sin\alpha-i\cos\alpha}}{\pi^{1/4}\sqrt{(\sin\alpha-i\sigma^2 \cos\alpha)}}\exp(\frac{-q^2}{2}\frac{\cot\alpha+i\sigma^2}{\sigma^2\cot\alpha+i}).
\label{tfgaussian}
\end{equation}
Note that the amplitude $F_\alpha(q)$ remains a real function and is invariant with $\alpha$ when $\sigma=1$. In the general  case,  it appears that the transform of a narrow function is wide and reverse, but it appears also that the $F_\alpha(q)$ function is not a Gaussian for $\alpha\in]0,\pi/2[$. Nevertheless the module of this function remains a Gaussian in agreement with the experiment.  The amplitude of the beam  has a width $\delta=\frac{1}{\sigma \sqrt 2}\sqrt{1-\cos(2\alpha)+\sigma^4(1+\cos(2\alpha))}$.

 \subsection{Fractional Fourier transform and the uncertainty principle}

In terms of quantum mechanics, every function $f(x)$ with a normalized probability density function($\int_{-\infty}^\infty |f(x)|^2dx=1$) and its Fourier transform $F(q)$ fulfilled the inequality:
\begin{equation}
{\rm Var}[f(x)]\times {\rm Var}[F(q)]\geq \frac14,
\end{equation}
with ${\rm Var}[f(x)]=\int_{-\infty}^\infty (x-\overline{x^2})^2 |f(x)|^2 dx$  and \\
$\overline{x}=\int_{-\infty}^\infty x |f(x)|^2 dx$.
This principle is a direct property of the standard Fourier transform. As respect to the FrFT, the uncertainty principle appears to be a peculiar case for $\alpha=\frac \pi 2$ and can be generalized for any $\alpha$:
 \begin{equation}
{\rm Var}[f(x)]\times {\rm Var}[\mathcal{F}^{\alpha}f(x)] \ \ \ \ \alpha\in[0,\pi/2]
\end{equation}

\subsubsection{The generalized uncertainty principle for a Gaussian beam}

Let's first consider the case of a Gaussian beam as given in equation \ref{gaussianbeam}
which fulfilled $\int_{-\infty}^\infty |f(x)|^2 dx=1$  and ${\rm Var}[f(x)]=\frac{\sigma^2}{2}$.
The Fourier transform F(q) gives: 
\begin{equation}
F(q)=\mathcal{F}_ \frac\pi 2 f(x)=\frac{\sqrt\sigma}{\pi^{\frac 14}}\exp({-\frac{\sigma^2 q^2}{2}})
\end{equation}
with ${\rm Var}[F(q)]=\frac {1}{2\sigma^2}$.
The minimum value is obtained for the Gaussian probability function:
\begin{equation}
{\rm Var}[f(x)]\times {\rm Var}[\mathcal{F}_  {\frac\pi 2}f(x)]= \frac{1}{4}.
\end{equation}
The variance of the Gaussian beam along the propagation for any angle $\alpha$ is given by the equation \ref{tfgaussian}:
\begin{equation}
{\rm Var}[\mathcal{F}_  {\alpha}f(x)]=\frac{\sigma^2 \cos^2\alpha}{2}+\frac{\sin^2\alpha}{2\sigma^2}.
\end{equation}
Let's consider the relation which gives the generalized incertitude principle from two applications of the operator on f(x) (with first $\mathcal{F}_  {\alpha}$ and then $\mathcal{F}_  {\beta}$) in the case of a Gaussian function and displayed in Fig~ \ref{uncertainty}b :
\begin{equation}
{\rm Var}[f(x)]\times {\rm Var}[\mathcal{F}_  {\alpha}f(x)]= \frac{\sigma^4 \cos^2\alpha}{4}+\frac{\sin^2\alpha}{4}
\end{equation}
and compare to that given by Shen \cite{shen} in the general case of any function $\phi$:
\begin{equation}
{\rm Var}[\phi]\times {\rm Var}[\mathcal{F}_  {\alpha}\phi]\geq\frac{\sin^2\alpha}{4}
\end{equation}

\subsubsection{The generalized uncertainty principle for a rectangular function}

The rectangular function $s(x)$ is a peculiar case. Because of the abrupt discontinuity,
the variance of  $\mathcal{F}_  {\alpha}s(x)$ is infinite whatever $\alpha \neq 0$.
The intensity profile in Fig~\ref{fresnel_fraunhofer} never vanishes completely, from the Fresnel to the Fraunhofer
regime\cite{infinite}. In terms of quantum mechanics, the probability for finding the particle anywhere until the first moment is not zero.

To treat diffraction of slit by considering a rectangular function is not completely right from an experimental point of view, since absorption through the blades of the slit induces a not abrupt truncation of the incident wave front\cite{lebolloch} and thus a finite variance. To take into account this effect, we could apply the FrFT to a rectangular function convoluted by a Gaussian function $g(x)$,
\begin{equation}
\mathcal{F}_  {\alpha}[s(x)\otimes g(x)],
\end{equation}
which would lead to intensity profiles with finite extension. 
Too large Gaussian functions smooth
the diffraction pattern and reduce the wave extension but may make the dark spot disappear.
It is difficult to sufficiently reduce the wave extension without vanishing the dark spot, which is observed experimentally.
 We  thus measure the variance over a limited area centered at the maximum intensity. For each $\alpha$, the full width at half maximum is measured and the {\it rms} variance is calculated over 2.35 times the FWHM. This is justify from an experimental point of view since, in most cases, the lack of intensity or finite sizes of detectors do not allow us to measure the diffraction pattern of slit far from the direct beam. The result is displayed in Fig~\ref{uncertainty}d).
 A clear minimum is obtained for $\alpha=\beta=\frac{\pi}{10}$ which shows that a successive coherent diffraction of two apertures focuses  beams in the Fresnel regime. This is the basic idea of  Fresnel zone plates used in X-rays.

\subsection{Fractional Fourier transform and diffraction of a periodic modulation}

\begin{figure}[!ht]
$$\resizebox{0.45\textwidth}{!}{%
\includegraphics{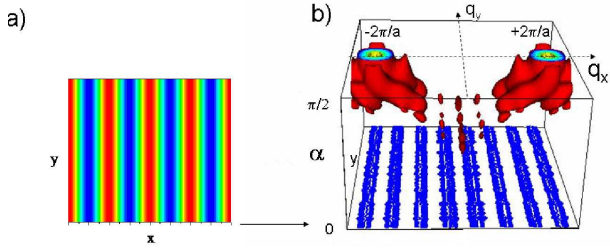}
}$$
\caption{a) 2D periodic modulation defined by a 1D wave vector $\vec q_0=\frac{2\pi}{a}\vec x$. b) Corresponding Fractional Fourier transform displayed for $\alpha\in]0,\frac \pi 2]$.  In the reciprocal space ($\alpha=\frac \pi 2$), the two Bragg reflections are located at $q=\pm q_0$. The fringes around $q=\pm q_0$ are due to the finite size object.}
\label{diffraction}
\end{figure}

Let's consider a 2D periodic modulation 
defined by a single wave vector $q_0=\frac{2\pi}{a}$, such as:
\begin{equation}
\rho(x,y)=\rho_0\cos(q_0 x),
\end{equation}
which gives rise  to two Bragg reflections  at $\pm q_0$ by diffraction. 
By continuously varying  $\alpha$ from $0$ to $\frac \pi 2$, the FrFT simply allows us to calculate the continuous evolution of the diffraction pattern  from the real space to the reciprocal space (see Fig~\ref{diffraction}).

\subsubsection{Successive diffraction of two objects}
As discussed in the introduction, the use of rectangular slits is necessary to obtain coherent X-ray beams from synchrotron sources
and their location relative to the diffracted object may influence diffraction patterns. To quantify this effect, the diffraction of two  successive objects has to be taken into account: diffraction of a rectangular function $s(x)$ followed by  the modulation $\rho(x)$,  as respect to $(\alpha,\beta)\in[0,\frac \pi 2]$:
\begin{equation}
\mathcal{F}_{\alpha}[\mathcal{F}_{\beta}[s(x)]\times\rho(x)].
\label{double_diff}
\end{equation} 

\begin{figure}[!ht]
$$\resizebox{0.45\textwidth}{!}{%
\includegraphics{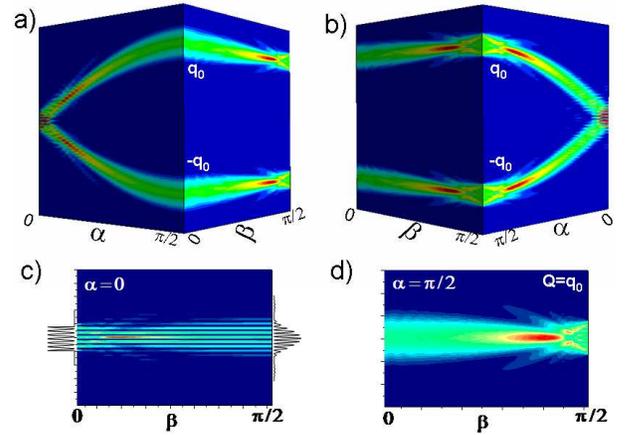}
}$$
\caption{a) and b) Successive diffraction of a rectangular function and a periodic modulation versus the slit-sample distance ($ \approx \beta$) and the sample-detector distance ($\approx \alpha$). c) for $\alpha=0$, the diffraction pattern corresponds to the Fig. 5c times $\cos (q_0 x)$  d) For $\alpha=\pi/2$, the Bragg reflection at q=$q_0$ displays a cardinal sinus squared profile for $\beta=0$ and a Fresnel diffraction profile when $\beta=\pi /2$.\label{inversion}}
\end{figure}

If the sample is located in the Fraunhofer regime of the aperture ($\alpha=\pi/2$) and the detector in the Fraunhofer regime of the sample ($\beta=\pi/2$):
\begin{equation}
\mathcal{F}_{\frac\pi 2}[\mathcal{F}_{\frac\pi 2}[s(x)]\times\rho(x)]=s(x)\otimes\mathcal{F}_{\frac\pi 2}[\rho(x)].
\end{equation}
The Bragg peak will mainly display the diffraction pattern of the rectangular function in the Fresnel regime.
If now the sample is located in the Fresnel regime of the aperture ($\beta\approx0$) and the detector in the Fraunhofer regime ($ \alpha=\frac\pi 2$), eq.~\ref{double_diff} gives:
\begin{equation}
\mathcal{F}_{\frac\pi 2}[\mathcal{F}_{0}[s(x)]\times\rho(x)]=\mathcal{F}_{\frac\pi 2}[s(x)]\otimes\mathcal{F}_{\frac\pi 2}[\rho(x)]. 
\end{equation}
In that case, the reflection profile is a convolution of the FT of the aperture with the FT of the periodic modulation.
The  profile versus the distance between the aperture and the sample is summarized Fig~\ref{inversion}. It is worthwhile to note that the Fig. 7d  corresponds to the inverse of the Fig~\ref{fresnel_fraunhofer}. By observing Bragg reflection, the double diffraction is similar to a time reversal operator.

\begin{figure}[!ht]
$$\resizebox{0.45\textwidth}{!}{%
\includegraphics{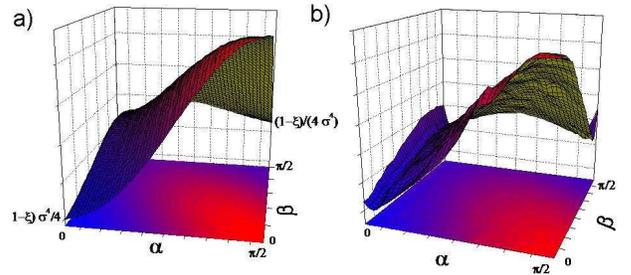}
}$$
\caption{ Product of variances in the case of the diffractions two objects: ${\rm Var}(\mathcal{F}_{\alpha}[\mathcal{F}_{\beta}[s(x)]\times\rho(x)])$ with $(\alpha,\beta)\in [0,\frac \pi 2]$  in the case of a) a Gaussian incident beam or b) in the case of a  rectangular function. The rms variance is calculated from the  Bragg peak maximum. With $\xi_1=2q^2\sigma^2\ (1+\exp[q^2\sigma^2])$ et $\xi_2=2/ (q^2\sigma^2) \ (1+\exp[q^2\sigma^2])$.\label{double_diff2}}
\end{figure}

The product of variances ${\rm Var}[\mathcal{F}_{\alpha} s(x)]$ times ${\rm Var}\mathcal{F}_{\frac\pi 2}[\mathcal{F}_{\alpha}[s(x)]\times\rho(x)]$ is displayed in Fig~\ref{double_diff2} in the case of a Gaussian beam and a rectangular aperture.
It is clear that to increase the width of Bragg reflection, the slit has to be as close as possible from the sample.

To conclude,  the fractional Fourier transform appears to be appropriate to treat coherent diffraction. Especially, the property of continuity of the FrFT could be useful for iterative reconstruction algorithms as ptychography\cite{dierof}.


\begin{thebibliography}{99}
\bibitem{infinite}In the Frauhofer regime, the variance of the cardinal sinus squared is infinite:
$\bar{x^2}=\int_{-\infty}^\infty x^2|f(x)|^2dx=\frac 2a\int_{-\infty}^{\infty} \sin^2(x\frac a2)dx$
\end{thebibliography}
\end{document}